\documentclass[a4paper,11pt]{article}
\pdfoutput=1 

\usepackage{jinstpub} 
\usepackage{color}

\title{\boldmath Estimation of reliable range of electron temperature measurements with sets of given optical bandpass filters for KSTAR Thomson scattering system based on synthetic Thomson data}

\author[a]{Keon-hee Kim,}
\author[a]{Tae-suk Oh,}
\author[a]{Kyeo-reh Park,}
\author[b]{J.H. Lee,}
\author[a, 1]{Y.-c. Ghim,\note{Corresponding author.}}

\affiliation[a]{Department of Nuclear and Quantum Engineering, KAIST, Daejeon, Korea}
\affiliation[b]{National Fusion Research Institute, Daejeon, Korea}

\emailAdd{ycghim@kaist.ac.kr}

\abstract{One factor determining the reliability of measurements of electron temperature using a Thomson scattering (TS) system is transmittance of the optical bandpass filters in polychromators. We investigate the system performance as a function of electron temperature to determine reliable range of measurements for a given set of the optical bandpass filters. We show that such a reliability, i.e., both bias and random errors, can be obtained by building a forward model of the KSTAR TS system to generate synthetic TS data with the prescribed electron temperature and density profiles. The prescribed profiles are compared with the estimated ones to quantify both bias and random errors.}


\proceeding{18$^{\text{th}}$ International Symposium on Laser-Aided Plasma Diagnostics\\
  24 - 28 September 2017\\
  Prague, Czech Republic}

\begin{document}
\maketitle
\flushbottom

\section{Introduction}\label{sec:intro}

\begin{figure}[b]
\centering 
\includegraphics[width=1.0\textwidth, origin=c]{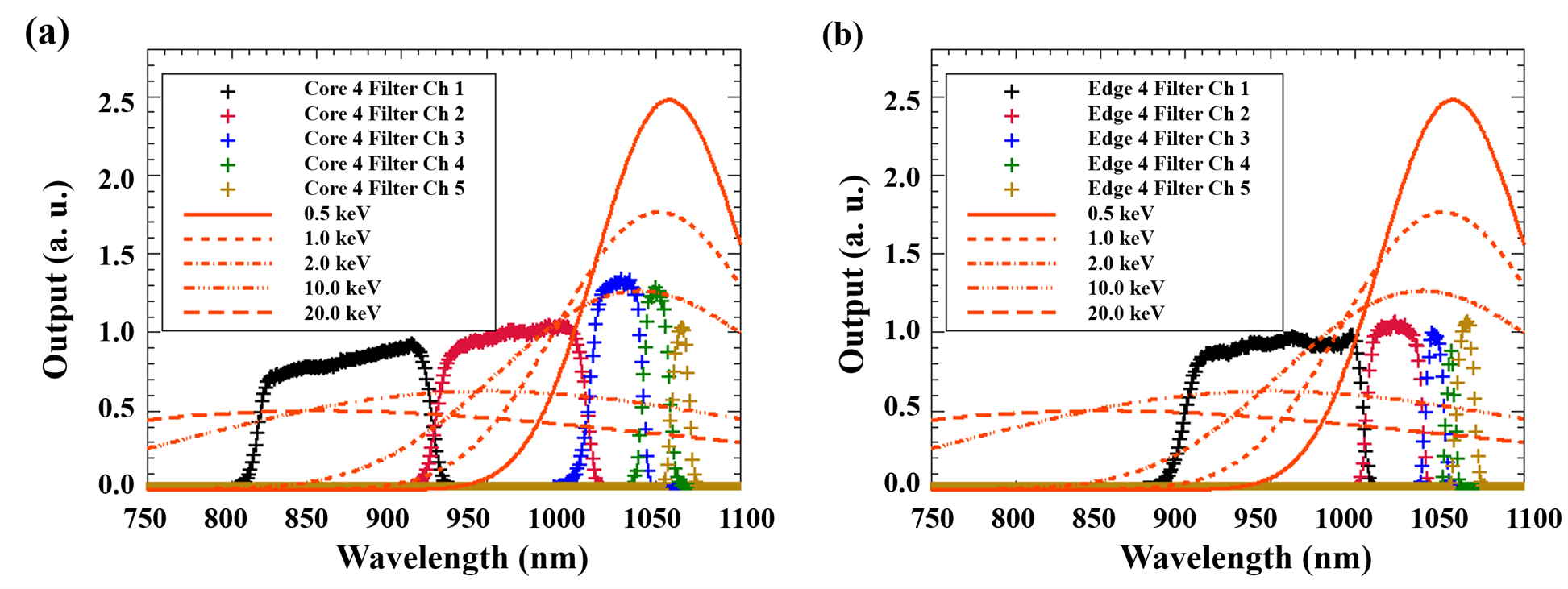}
\caption{\label{fig:filter} Examples of measured transmittance of the optical bandpass filters (+) with spectral distributions of Thomson scattered lights for various electron temperatures (lines) for (a) Core \#4 and (b) Edge \#4 of the KSTAR TS polychromators used during the 2016 Campaign.}
\end{figure}

Thomson scattering (TS) diagnostic system is widely used to measure both temperature and density of electrons in hot fusion-grade plasmas. Broadening of the spectral distribution of the Thomson scattered light is a function of electron temperature; whereas the intensity of the light contains the information of electron density~\cite{Naito}. The KSTART TS system collects the scattered light with polychromators consisting of a set of five optical bandpass filters~\cite{JHLee} as shown in figure~\ref{fig:filter}. Reliably measurable ranges of temperature depend on the properties of the filters such as central wavelength, bandwidth and the number of the filters. 

Achievable fractional errors given a set of the filters together with TS system parameters have been estimated by considering measured signal level of background and scattered light with an electronic, i.e., amplifiers, noise~\cite{Scannell_thesis, Scannell, HGLee}. Such fractional errors provide quantitative magnitude of random errors but do not present bias errors. With the aim of obtaining both bias (if finite) and random errors, we have developed a forward model of the KSTAR TS system and generated synthetic TS data~\cite{TSOh} with prescribed profiles of electron density and temperature. The forward model is based on the TS system parameters used during the 2016 KSTAR campaign.

\begin{figure}[t]
\centering
\includegraphics[width=0.8\textwidth, origin=c]{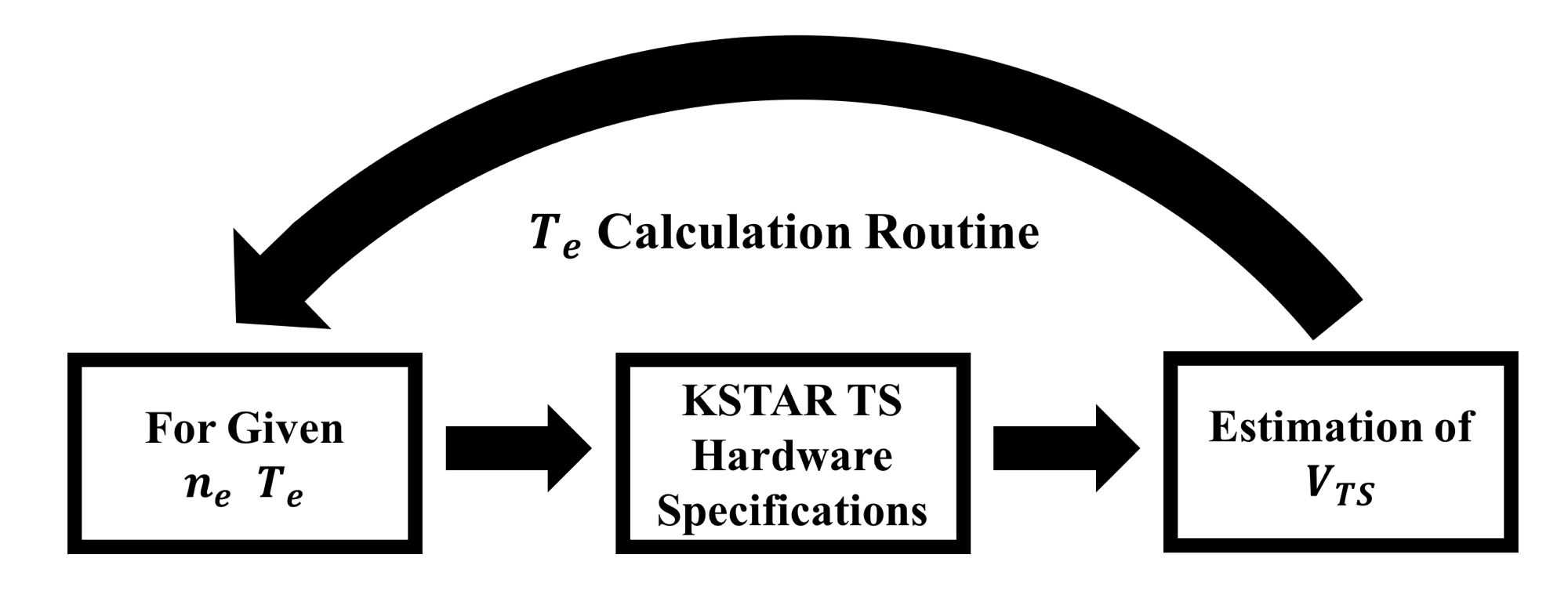}
\caption{\label{fig:forward} A schematic of the KSTAR TS forward model. Expected (synthetic) TS data are generated based on the KSTAR TS system specifications, and the synthetic data are analyzed as if they were actual experimental data. Then, results are compared with the prescribed profiles of electron density and temperature to estimate the reliability of the KSTAR TS system.}
\end{figure}

Treating the synthetic TS data as if they were experimentally measured ones, we can obtain temperature and density of electrons by using a look-up table method~\cite{TSOh, SOh} which, then, are compared with the prescribed temperature and density. Figure~\ref{fig:forward} schematically describes such a procedure. This will allows us to quantify both bias and random errors. In this paper, we investigate reliability of estimating various profiles of electron temperature with a fixed density profile.

\section{Forward model of the KSTAR Thomson scattering system}\label{sec:forward_model}

\begin{figure}[t]
\centering
\includegraphics[width=0.5\textwidth, origin=c]{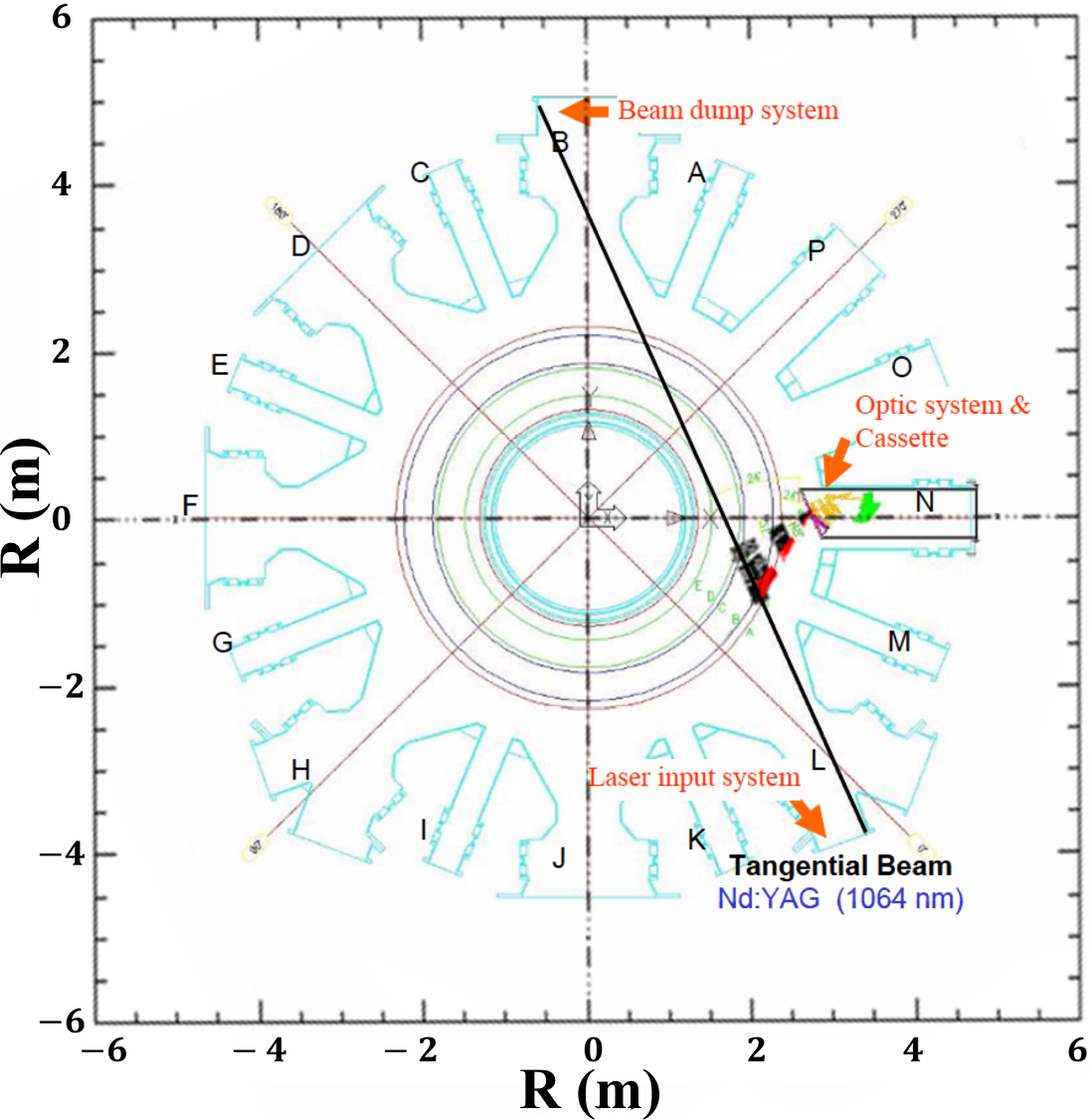}
\caption{\label{fig:geometry} Top view of the KSTAR TS system. Nd:YAG laser pulse with the wavelength of $1064$ nm is injected tangentially from the L-port, and the laser beam dump is located at the B-port. Thomson scattered light is collected via edge (red dashed lines) and core (black dashed lines) optical systems at the N-port.}
\end{figure}

Laser pulse with the wavelength of $1064$ nm is tangentially injected from the L-port of KSTAR~\cite{JHLee_2012} and the Thomson scattered photons are collected via edge and core optical systems as shown in  figure~\ref{fig:geometry}. These photons are passed to polychromators, where each polychromator contains five optical bandpass filters. Combinations of central wavelengths and bandwidths are different between core and edge polychromators as shown in figure~\ref{fig:filter}(a) and (b), respectively. Photons passed through a bandpass filter are detected by an avalanche photodiode detector (APD: Hamamatsu S11519-30). Since there are five bandpass filters in a polychromator, we have five APDs for each polychromator. 

Each APD outputs an electronic signal proportional to the detected photons, and this signal can be modelled as~\cite{Scannell_thesis, TSOh}, 
\begin{equation}
\label{eq:Thomson_signal}
V_{\text{TS}}^i=G~n_e~N_\text{laser}~\frac{d\sigma_\text{TS}}{d\Omega} \Delta\Omega~L~T(\lambda_\text{L})~QE
~\int~\frac{\phi^i(\lambda_\text{s})}{\phi(\lambda_\text{L})}\frac{S(\lambda_\text{S}; T_e, \theta, \lambda_\text{L})}{\lambda_\text{L}}~d\lambda_\text{s},
\end{equation}
where superscript $i$ denotes the $i^\text{th}$ channel of the optical bandpass filters in a given polychromator. $G$, $n_e$, $N_\text{laser}$ and $\frac{d\sigma_\text{TS}}{d\Omega}$ correspond to the APD gain factor, the electron density, the number of injected photons in a single laser pulse and the differential Thomson scattering cross-section, respectively.  $\Delta\Omega$ and $L$ are the the solid angle and the scattering length of the KSTAR TS system, respectively. Transmission coefficient is a function of wavelength, and this is captured via taking the absolute coefficient at the laser wavelength $T(\lambda_\text{L})$ ($\lambda_\text{L}=1064$ nm is the laser wavelength) with the normalized filter transmittance function (or just simply filter function) $\phi^i(\lambda_\text{s})/\phi(\lambda_\text{L})$, where $\lambda_\text{s}$ is the scattered wavelength. Figure~\ref{fig:filter}(a) and (b) show such normalized filter functions (+). Note that the quantum efficiency $QE$  of the APD detector is also a function of the wavelength, but such a variation is captured by the filter function $\phi^i(\lambda_\text{s})$ as well; thus $QE$ in equation~(\ref{eq:Thomson_signal}) is the quantum efficiency at the laser wavelength. $S(\lambda_\text{S}; T_e, \theta, \lambda_\text{L})$ is a spectral distribution of the Thomson scattered light~\cite{Naito, Prunty} as a function of the scattered wavelength $\lambda_\text{s}$ at a certain electron temperature $T_e$, scattering angle $\theta$ and laser wavelength $\lambda_\text{L}$.

\begin{table}[b]
\centering
\caption{\label{table:hardware_spec} Specification of the KSTAR TS system used during the 2016 Campaign.}
\smallskip
\begin{tabular}{|l|cc|}
\hline
&Core Optical System&Edge Optical System\\
\hline\hline
Laser Energy&$2$~J&$2$~J\\
F/\#&$5.6$&$6.5$\\
Solid Angle ($\Delta\Omega$)&0.0250&0.0186\\
Transmission Coefficient at $\lambda_\text{L}$ ($T$)&0.57&0.57\\
Quantum Efficiency $\lambda_\text{L}$ ($QE$)&0.58&0.58\\
\hline
\end{tabular}
\end{table}

\begin{figure}[t]
\centering
\includegraphics[width=0.95\textwidth, origin=c]{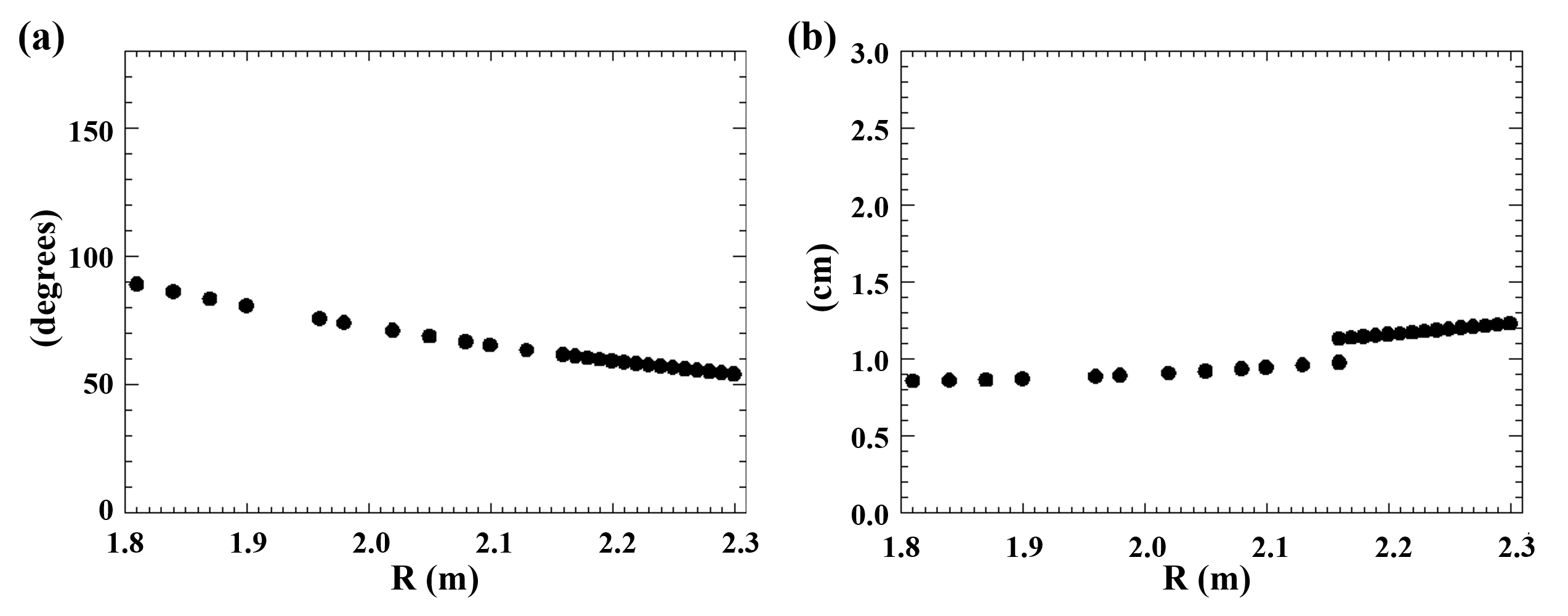}
\caption{\label{fig:angle_length} Calculated (a) scattering angle $\theta$ and (b) scattering length $L$ as a function of the major radius $R$ for the KSTAR TS system used during the 2016 Campaign.}
\end{figure}

To be able to generate expected (or synthetic) KSTAR TS data based on equation~(\ref{eq:Thomson_signal}), we need to use hardware specifications of the KSTAR TS system which are listed in table~\ref{table:hardware_spec}. The KSTAR TS system measures electron density and temperature at 12 spatial positions with the core optical system, i.e., $R=1.81$, $1.84$, $1.87$, $1.90$, $1.96$, $1.98$, $2.02$, $2.05$, $2.08$, $2.10$, $2.13$ and $2.16$~m, and 15 spatial positions with the edge optical system, i.e., $R=2.16$, $2.17$, $2.18$, $2.19$, $2.20$, $2.21$, $2.22$, $2.23$, $2.24$, $2.25$, $2.26$, $2.27$, $2.28$, $2.29$ and $2.30$~m, where $R$ is the major radius. Note that a typical location of the magnetic axis in KSTAR is $R_0\approx1.80$~m. Depending on the spatial positions, scattering angle $\theta$ and scattering length $L$ are different, and they have been calculated based on the geometry of the KSTAR TS system (e.g. figure~\ref{fig:geometry}). Scattering angle and length as a function of the major radius $R$ are shown in figure~\ref{fig:angle_length}.

Collected number of scattered photons as a function of the scattered wavelength denoted as $n_\text{s}$ just before passing a polychromator can be written as
\begin{equation}
\label{eq:num_ph}
n_\text{s}(\lambda_\text{s})=n_e~N_\text{laser}~\frac{d\sigma_\text{TS}}{d\Omega} \Delta\Omega~L~T(\lambda_\text{L})
~\frac{S(\lambda_\text{S}; T_e, \theta, \lambda_\text{L})}{\lambda_\text{L}},
\end{equation}
and the number of photo-electrons $N_\text{s}$ from an APD detector can be written as
\begin{equation}
\label{eq:num_ph}
N_\text{s}(\lambda_\text{s})= n_\text{s}(\lambda_\text{s})~QE~\frac{\phi^i(\lambda_\text{s})}{\phi(\lambda_\text{L})}.
\end{equation}
\begin{figure}[b]
\centering
\centering
\includegraphics[width=.5\linewidth]{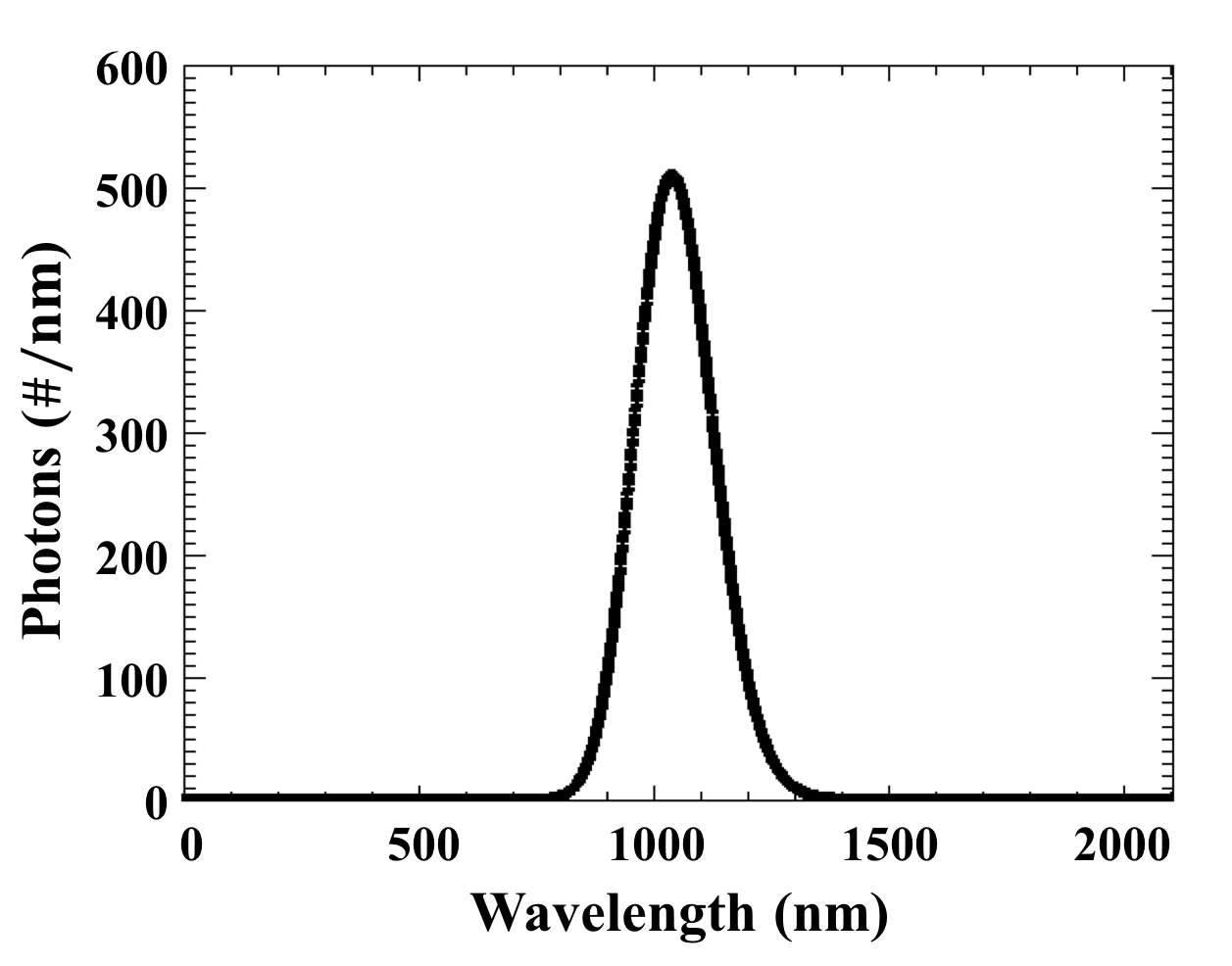}
\caption{\label{fig:num_ph} Spectral distribution of collected scattered photons $n_\text{s}(\lambda_\text{s})$ for Core \#4 ($R=1.90$~m) when $T_e=1.9$~keV and $n_e=1.0\times10^{19}$~m$^{-3}$.}
\end{figure}
\begin{figure}[t]
\centering
\includegraphics[width=.5\linewidth]{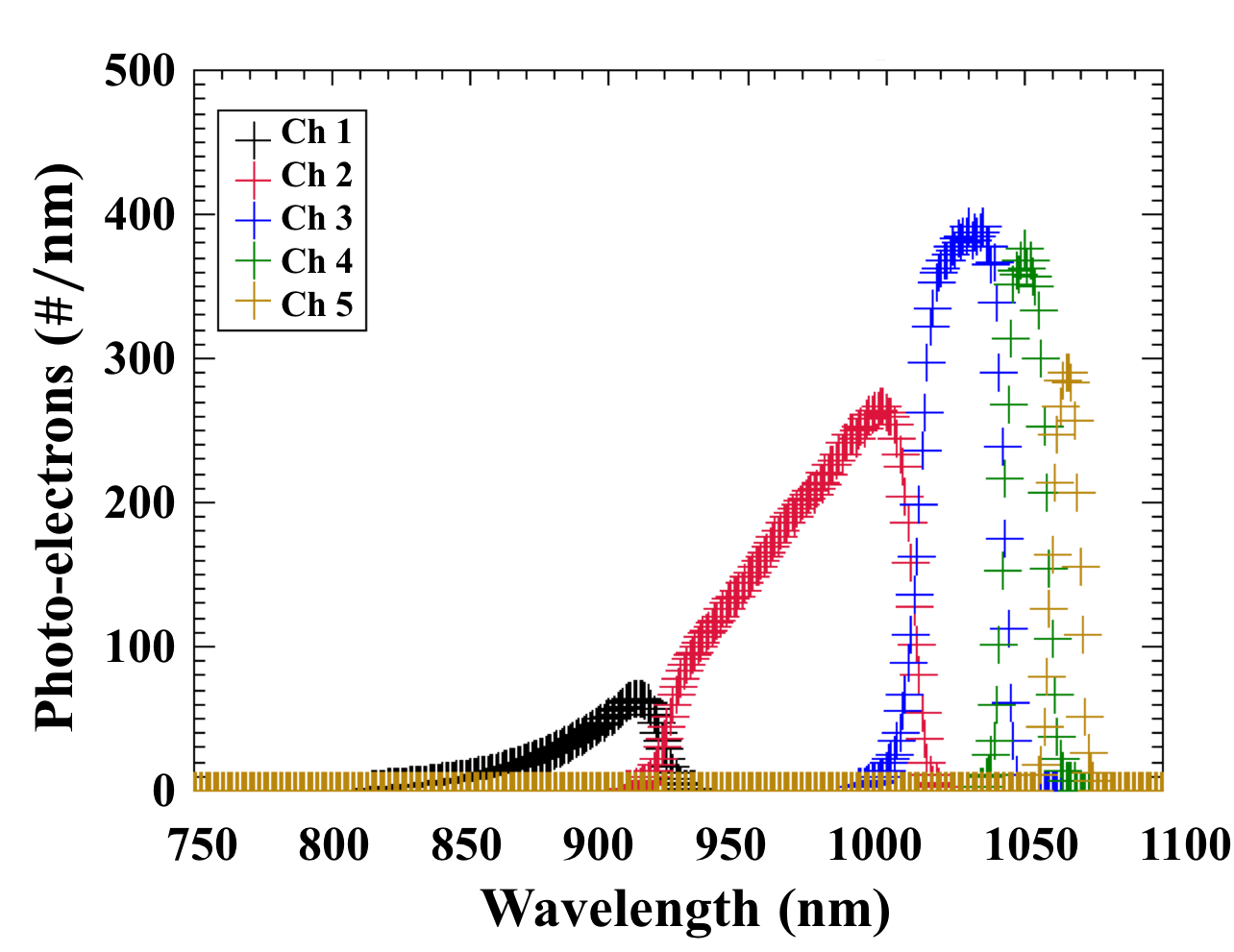}
\caption{\label{fig:num_phe} Spectral distribution of photo-electrons $N_\text{s}(\lambda_\text{s})$ detected by five APD detectors of the Core \#4 polychromator ($R=1.90$~m) when $T_e=1.9$~keV and $n_e=1.0\times10^{19}$~m$^{-3}$.}
\end{figure}
Figures~\ref{fig:num_ph} and \ref{fig:num_phe} show examples of collected scattered photons $n_\text{s}(\lambda_\text{s})$ and photo-electrons $N_\text{s}(\lambda_\text{s})$ detected by the Core \#4 polychromator ($R=1.90$~m) when $T_e=1.9$~keV and $n_e=1.0\times10^{19}$~m$^{-3}$. Integrating photo-electrons with respect to the scattered wavelength $\lambda_\text{s}$ and multiplying it with the APD gain factor $G$ provides the electronic signal that we can read using a digitizer, i.e., equation~(\ref{eq:Thomson_signal}). For the purpose of this paper, actual value of $G$ is irrelevant because the KSTAR TS system is dominated by the photon noise~\cite{TSOh} which means that noise levels are determined by the number of photo-electrons. We note that our forward model can act as a likelihood function in the frame of Bayesian probability theory~\cite{Sivia}.

\section{Comparison between the prescribed $T_e$ and the estimated $T_e$}\label{sec:comparison}

\begin{figure}[t]
\centering
\includegraphics[width=.5\linewidth]{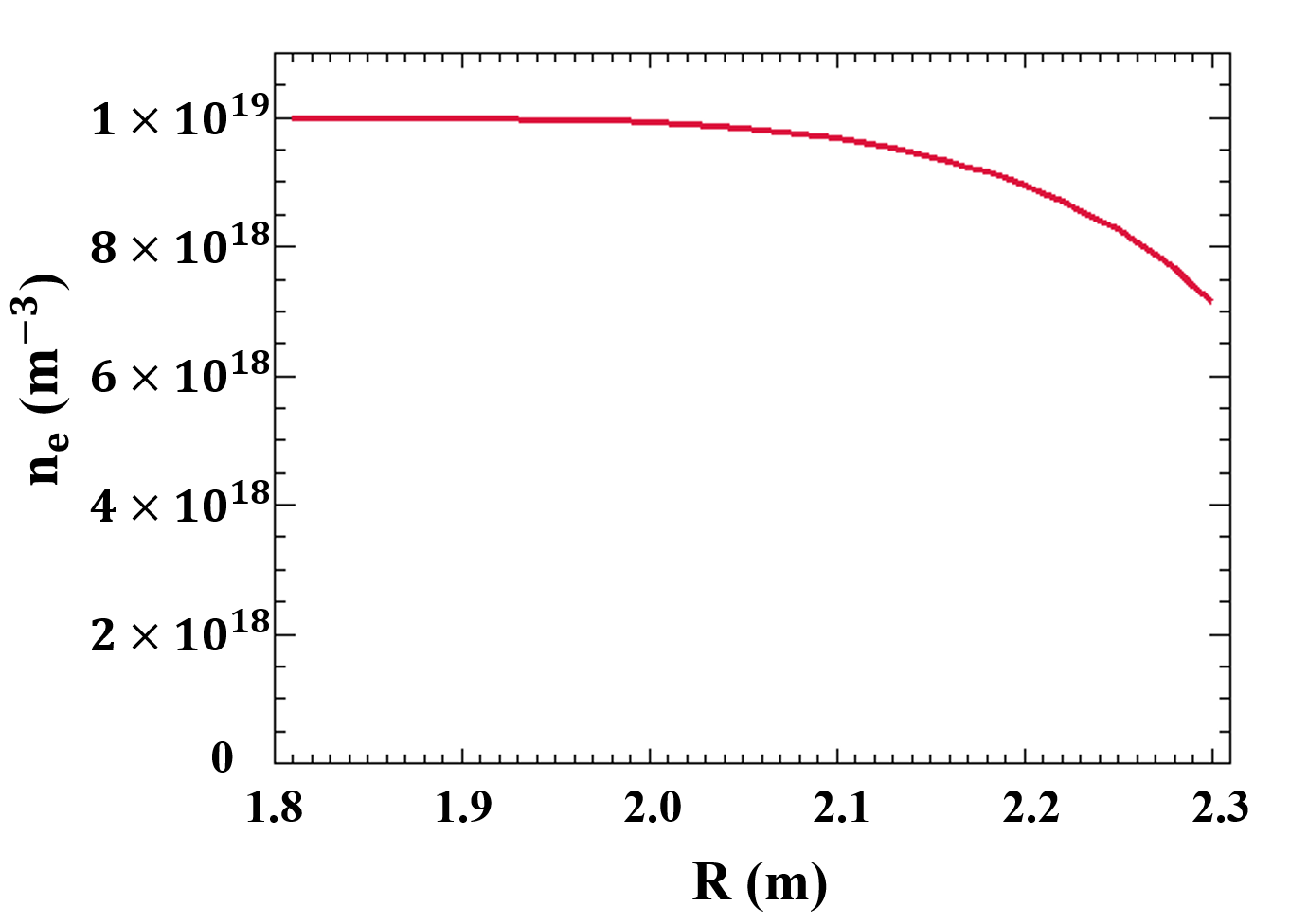}
\caption{\label{fig:dens_profile} A typical electron density profile of L-mode discharge observed in KSTAR.}
\end{figure}

\begin{figure}[t]
\centering
\includegraphics[width=.7\linewidth]{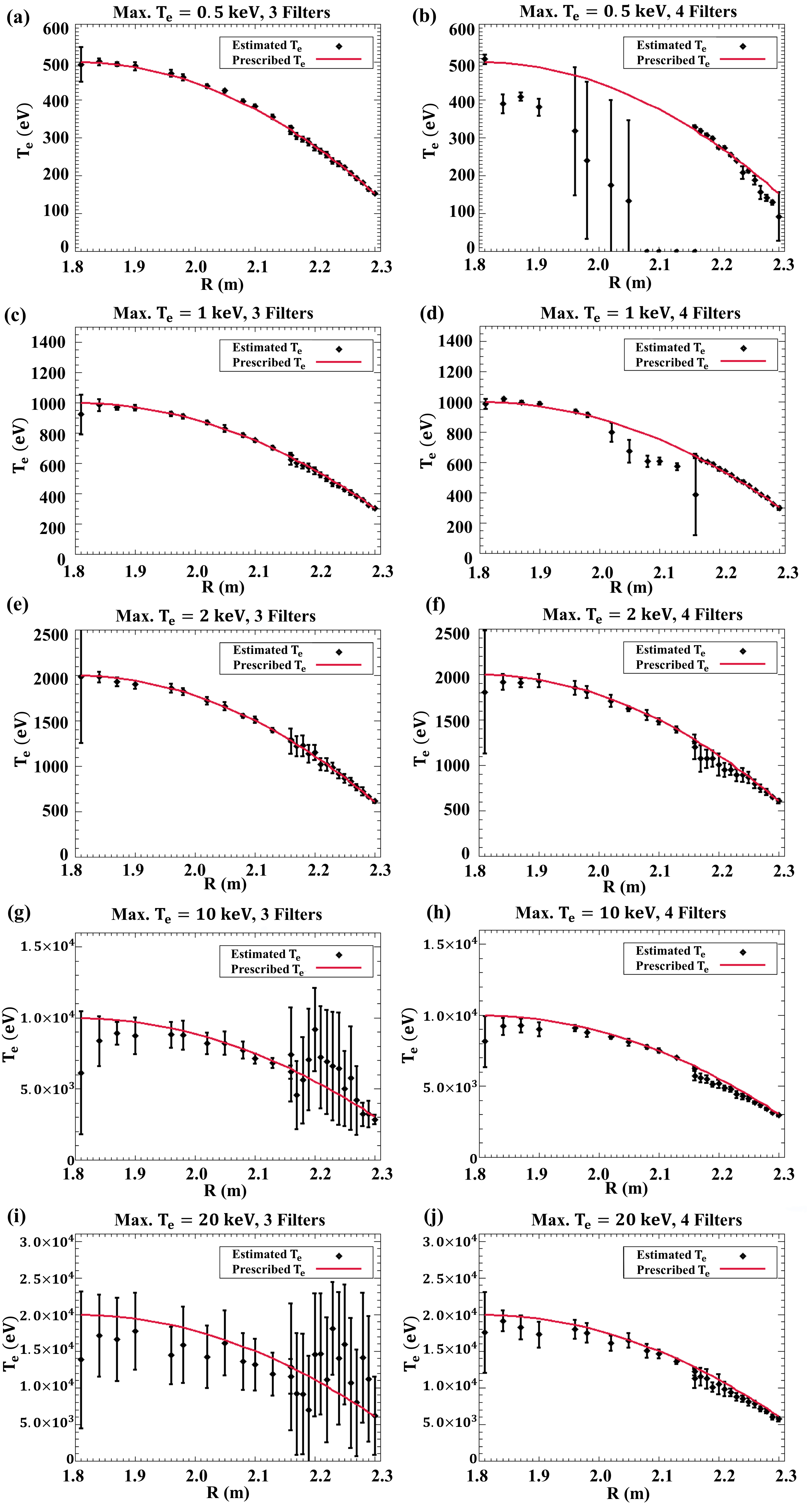}
\caption{\label{fig:temp_profile} Comparison between the prescribed (red line) and estimated (dots) electron temperature $T_e$ based on the KSTAR TS synthetic data. Ten data sets are used to estimate a profile of $T_e$ for each prescribed $T_e$ profile. Means are marked with dots, whereas error bars represent standard deviations.}
\end{figure}

Based on the KSTAR TS forward model described in section~\ref{sec:forward_model}, we generate the KSTAR TS synthetic data with a typical electron density profile of L-mode discharge observed in KSTAR as shown in figure~\ref{fig:dens_profile}, while varying profiles of electron temperature. Depending on the total number of photo-electrons from an APD detector (e.g. figure~\ref{fig:num_phe}), we add random Poisson noise to the signal. Using a look-up table method~\cite{TSOh, SOh}, we estimate electron temperature from the noise added synthetic data. Although the KSTAR TS system contains five bandpass filters and five APD detectors denoted as Ch.1 to Ch.5 for each polychromator, Ch.5 is excluded from the temperature estimation because Ch.5 is strongly affected by the stray light~\cite{TSOh} as the laser wavelength, i.e., $1064$~nm, is within the bandpass of this channel (see figure~\ref{fig:filter}). In using the look-up table method, we have examined two cases: 1) using three filters (or channels), i.e., Ch.2, Ch.3 and Ch.4, and 2) using four filters (or channels), i.e., Ch.1, Ch.2, Ch.3 and Ch.4. We generate ten data sets from which mean and random error (standard deviation) are estimated for a given temperature profile where the randomness is added by the Poisson noise. Comparisons between the prescribed and estimated electron temperatures for these two cases with various electron temperatures are shown in figure~\ref{fig:temp_profile}. Note that the polychromator observing the most inner location, i.e., $R=1.81$~m uses `edge' bandpass filters even if it is defined as `core' polychromators.

For the core polychromator system whose combination of bandpass filters is shown in figure~\ref{fig:filter}(a), we find that using four filters (Ch.1 to Ch.4) for estimating $T_e < 0.8$~keV results in underestimation, i.e., bias error, and relatively larger uncertainties compared to using three filters (Ch.2 to Ch.4) from figure~\ref{fig:temp_profile}(a)-(d). Since the bandpass of Ch.1 is farthest away from the laser wavelength, relatively low temperature generates very small amount of photo-electrons for Ch.1 resulting in a very small signal-to-noise ratio. On the other hand, using four channels gives us better temperature estimation for $T_e > 10$~keV as in figure~\ref{fig:temp_profile}(g)-(j). For such a large temperature, we now have a good signal-to-noise ratio for Ch.1, and this channel starts to provide additional valuable information for estimating temperature. 

For the edge polychromator system shown in figure~\ref{fig:filter}(b), we find that as long as $T_e < 5$~keV, either using three filters or four filters provides us reasonably good estimation of temperatures. If $T_e > 5$~keV, then using three filters results in the overestimation and larger uncertainties for estimating temperature compared to using four channels as in figure~\ref{fig:temp_profile}(g)-(j).

\section{Conclusion}\label{sec:conclusion} 

We have developed a forward model of the KSTAR TS system for the purpose of quantifying both bias and random errors in estimating electron temperature, where previous estimation of fractional errors provide just random errors. Our model can be used not only for the KSTAR TS system but also any other TS systems as the forward model is generic. For the examined KSTAR TS system, we find that three filters are to be used for core polychromators when electron temperature is less than $0.8$~keV, while using four filters are recommended when the temperature is larger than $10$~keV.  For the edge polychromators, choice of using three or four filters is irrelevant as long as the temperature is less than $5$~keV, and using four filters is recommended for the temperature greater than $5$~keV.

\acknowledgments
This work is supported by National R\&D Program through the National Research Foundation of Korea (NRF) funded by the Ministry of Science and ICT (grant numbers NRF-2017M1A7A1A01015892 and NRF-2017R1C1B2006248)  and the KUSTAR-KAIST Institute, KAIST, Korea.


\end{document}